Effect of a small disruption of the Ca site on the geometrically frustrated magnetic behavior of $Ca_3CoRhO_6$

Niharika Mohapatra and E.V. Sampathkumaran[*]

*Tata Institute of Fundamental Research, Homi Bhabha Road, Colaba, Mumbai – 400005, India.*

*Abstract*

The compound, $Ca_3CoRhO_6$, containing magnetic Co-Rh chains intervened by Ca ions, has been known to be one of the few exhibiting 'partially disordered antiferromagnetic structure (PDA)' due to geometrical frustration. Here, we report the influence of partial replacement of Ca by Y on the magnetic anomalies by investigating the solid solution, $Ca_{3-x}Y_xCoRhO_6$, by bulk measurements. There are profound changes in the magnetic behavior, the most notable one being that the features attributable to spin-chain magnetism and PDA structure get suppressed dramatically by a small replacement of Ca by Y $(x < 0.3)$, despite the fact that the magnetic chain is not disrupted. This finding suggests that this compound is on the verge of PDA-structural-instability.

In the field of 'geometrically frustrated magnetism', a few compounds exhibiting 'partially disordered antiferromagnetic structure (PDA) viz., $CsCoCl_3$ and $CsCoBr_3$, have been of great interest since seventies [1]. In the PDA structure, two of three magnetic chains, arranged in a triangular fashion in the basal plane, order antiferromagnetically with the 'third' chain remaining incoherent due to interchain antiferromagnetic coupling. In this respect, the compound, $Ca_3Co_2O_6$, crystallizing in the $K_4CdCl_6$-type rhombohedral structure (space group: $R\bar{3}c$), has been attracting a lot of attention in recent years [2]. In this compound, PDA structure forms in an intermediate temperature range below ($T_N=$ ) 30 K followed by a complex or frozen-PDA (F-PDA) structure due to freezing of the third chain below about ($T_f=$ ) 10 K [See, for example, Ref. 2 and articles cited therein]. Despite the fact that the compound, $Ca_3CoRhO_6$, has also been shown to exhibit similar magnetic structure with the corresponding temperatures being 90 and 40 K respectively [3-5], there is not much activity in this compound, barring few reports [6-9]. In this compound, Co occupying the trigonal prismatic site is in high-spin ($3d^7$) 2+ state, whereas Rh with octahedral coordination is in low-spin ($3d^5$) 4+ state. There is a broad hump in the temperature (T) dependence of magnetic susceptibility ($\chi$) in the range 90-150 K characteristic of one-dimensional short-range magnetic ordering as established by neutron diffraction data [7]. Thus, this compound apparently offers a rare opportunity to probe both spin-chain magnetism and geometrical frustration phenomena in a single system. Another interesting property of this compound is that the frequency ($\nu$) dependence of ac $\chi$ in the vicinity of magnetic transition is huge, typical of superparamagnets but not of stoichiometric compounds [6]. The aim of this work is to explore how the interesting magnetic characteristics of this compound are modified by disrupting the Ca site, intervening the magnetic chains, by Y substitution. It may be stated that this material may gain importance from the angle of thermoelectric applications as well, considering that the thermopower is very large [10].

The polycrystalline samples, $Ca_{3-x}Y_xCoRhO_6$ (x= 0.0, 0.15, 0.3, 0.5, 0.75 and 1.0), were prepared by a conventional solid state route as described in Ref. 6 starting from stoichiometric amounts of high purity (>99.9%) $CaCO_3$, $Y_2O_3$, CoO and Rh powder. The samples were characterized by x-ray diffraction (Cu $K_\alpha$) and scanning electron microscope (SEM). X-ray diffraction patterns are shown in figure 1. The single phase with composition homogeneity forms up to $x=$ 0.5 and therefore we present the data for these compositions only. We do not find any change in the oxygen content within the sensitivity limit (few percent) of SEM. The dc magnetization (M) measurements as a function of temperature (1.8-300 K) were carried out employing a commercial superconducting quantum interference device (SQUID) (Quantum Design, USA) for zero-field-cooled (ZFC) and field-cooled (FC) conditions of the specimens in the presence of (H= ) 100 Oe and 5 kOe; in addition, isothermal M data were taken at selected temperatures with a vibrating sample magnetometer (VSM, Oxford Instruments, UK). The same SQUID magnetometer was employed for ac $\chi$ measurements with different $\nu$.

The results of dc M measurements as a function of T are shown in figure 2. For the parent compound, as known in the literature [3,6], $\chi(T)$ is nearly flat in the range 90-150 K; as a signature of PDA regime, there is a sharper increase of M/H below about 90 K and the M/H values are H-dependent in the range 40-90 K; there is a peak and a sharp fall in ZFC-$\chi(T)$ at $T_f$ (=40 K) (which is absent in FC-curve), signaling the onset of a spin-glass-like phase. One of the major points of emphasis is that the above-mentioned PDA and spin-chain characteristics in magnetization are essentially suppressed even for a small replacement of Ca by Y, say for $x= 0.15$; that is, M/H increases rather smoothly with deceasing temperature without flattening or without any H-dependence till $T_f$ where ZFC-FC curves bifurcate. In addition, $T_f$, which is marked by a peak in ZFC-curve, undergoes a sharp decrease with this Y substitution, that is, from 40 K for $x= 0.0$ to 31K for $x= 0.15$. For further replacement of Ca, say for $x= 0.3$ and 0.5, the corresponding values are 29 and 23 K. At this juncture, we would like to mention that the heat-capacity decreases with temperature monotonically (not shown here) across $T_f$ for all compositions, which is consistent with random spin-freezing. Another noteworthy finding is that, below $T_f$, while the M/H values tend to a small constant value, the magnitude keeps increasing with increasing $x$. An additional shoulder (peak) appears in ZFC curve, but not in FC curves, in the 5-10 K range for $x= 0.3$ (0.5) (see figure 2), possibly indicating more complex magnetic behavior at low temperatures with gradual Y substitution. With respect to the paramagnetic state, the plots of inverse $\chi(T)$ in the paramagnetic state are found to be non-linear (Figure 3) for all compositions as for x= 0.0 [6] possibly due to growing influence of intrachain short range magnetic order with decreasing temperature; hence it is rather difficult to obtain reliable estimates [11] of the paramagnetic Curie temperature ($\theta_p$) and the effective moment ($\mu_{eff}$). However, an inference could be drawn qualitatively on the trends from a very narrow (above 225 K) high temperature Curie-Weiss region. $\theta_p$ thus obtained decreases with increasing $x$ (~140, 122, 106 and 78 K for x= 0.0, 0.15, 0.3 and 0.5 respectively), thereby suggesting that intra-chain ferromagnetic correlation strength is diminished by Y substitution. This could be due to the fact that $c$-axis expands with gradual replacement of Ca by Y; interchain antiferromagnetic interaction strength may also grow with the gradual reduction of $a$. With respect to the trends in the effective magnetic moment, the value obtained from the above Curie-Weiss region decreases marginally with $x$ (from ~5.1 $\mu_B$/f.u. for x= 0.0 to 4.8 $\mu_B$/f.u. for x= 0.5). Therefore, a small change in the valence/spin state of Co or Rh, with a different local chemical environment due to Y substitution, can not be ruled out. In this respect, this solid solution differs from the behavior of $Ca_{3-x}Y_xCo_2O_6$ [Ref. 12].

We now compare (Fig. 4) isothermal M studies at 60 (PDA region for x= 0.0) and 5 K (F-PDA region). At 60 K, for the parent compound, as known earlier [5,6], there is a step in M near 40-60 kOe due to the alignment of incoherent magnetic chains as a signature of zero-field PDA structure. This step vanishes even for $x= 0.15$ and M increases rather smoothly for all compositions till the highest field (120 kOe) employed. This endorses the conclusion that PDA structure is absent for all other compositions, $x \geq 0.15$. With respect to the behavior at 5 K, M is hysteretic and does not saturate even for an application of a field of 120 kOe for all compositions, typical of spin-glasses. For x= 0.5, we have taken the data at 15 K as well (not shown here) to understand possible changes in spin structure across the range 10-15 K as indicated by the appearance of a broad peak in ZFC-curve (see figure 2) and we do not find any qualitative change with respect to the M(H) curve of 5 K. A point of note in the 5K-data is that the magnitude of M at higher fields dramatically increases with $x$, $e.g.,$ from about 0.2 $\mu_B$/f.u. for x= 0.0 to about 1.6 $\mu_B$/f.u. for $x= 0.5$ at H= 120 kOe, unlike in $Ca_{3-x}Y_xCo_2O_6$ [Ref. 12], indicating a change in the valence/spin behavior of Co and Rh with substitution. In this respect, it is of interest to carry out spectroscopic studies on this solid solution.

With respect to ac $\chi$ behavior (see figure 5), we have taken data at four frequencies (1.3, 13, 133 and 1339 Hz). As presented in Ref. 6, the parent compound is characterized by a well-defined feature in the real ($\chi'$) part at 50 K and the peak shifts significantly with increasing $\nu$ to higher temperatures, e.g., by about 20 K as for $\nu$ is increased from 1.3 Hz to 1.339 kHz.; similar behavior is seen in the imaginary part ($\chi''$). In the presence of a magnetic field of 30 kOe (see figure 2 of Ref. 6), the curves recorded at different $\nu$ merge resulting in a broad peak in $\chi'$, but no feature could be observed in $\chi''$. With Y substitution, the peaks shift to a lower temperature; thus in $\chi'$, the peak appears at 40 K, 37 K and 30 K for $\nu$= 1.3 Hz for $x$= 0.15, 0.3 and 0.5 respectively, with a large $\nu$-dependence similar to x= 0.0 (with corresponding peaks in $\chi''$). If the data is taken in the presence of 30 kOe, the peaks still persist *without any significant change in $\nu$-dependence* [13]; in fact, interestingly, the peaks appears in $\chi''$ as well, though noisy in nature with increasing $x$; in these respects, ac $\chi$ behavior of the Y-substituted systems are distinctly different from that of parent compound. It is interesting that the features are insensitive to the application of even large magnetic fields after partial Y substitution (unlike in $x$= 0.0), unlike in conventional spin-glasses. These observations reveal complexities of spin dynamics in this solid solution.

We have also performed isothermal remnant magnetization ($M_{IRM}$) measurements at selected temperatures (in the two temperature regions of PDA and F-PDA), as described in Ref. 6. It was found that the value of $M_{IRM}$ falls to nearly zero value within few seconds after the field is switched off, not only in the paramagnetic state, but also at 60 K for all compositions except for parent compound. On the other hand, below $T_f$, say at 5 K, $M_{IRM}$, is significant large decaying slowly (logarithmically with the form, $M_{IRM}(t) = M_{IRM}(0)[1 - s \, lnt]$) with time, $t$, as for x= 0.0 (see figure 4, inset), thereby supporting glassy state of magnetic structure at very low temperatures. Though the values of $M_{IRM}$ at $t$= 0 increase with increasing $x$ (0.0117, 0.026, 0.159 and 0.18 emu/mol respectively), the coefficient $s$ interestingly remains nearly constant (~ 0.13), as though the relaxation rate in the very low temperature glassy state does not change with substitution.

Summarizing, replacement of a small fraction of Ca ions by Y ions in $Ca_3CoRhO_6$ causes pronounced changes in its magnetic behavior, apparently destroying PDA behavior. Partial substitution of Y for Ca stabilizes lower-temperature glassy behavior only, with a complex spin dynamics which is highly robust to the application of moderate magnetic fields, as indicated by ac $\chi$ data. It is generally difficult to get a reliable estimate of intrachain and interchain interaction strengths in this class of compounds [14]. However, the fact that a small change in $c$ and $a$ in the opposite direction is sufficient enough to kill PDA anomalies indicates that there could be a sensitive balance between these couplings playing a crucial role to decide this novel magnetic structure. Neutron diffraction, magnetic circular dichroism and photoemission studies as a function of temperature to understand valence/spin behavior of Co and Rh ions in this solid solution will be quite rewarding. Finally, we have also performed electrical resistivity ($\rho$) measurements and found that all compositions are insulators.

We thank Kartik K Iyer for his help in performing experiments.

References:
*E-mail: sampath@tifr.res.in

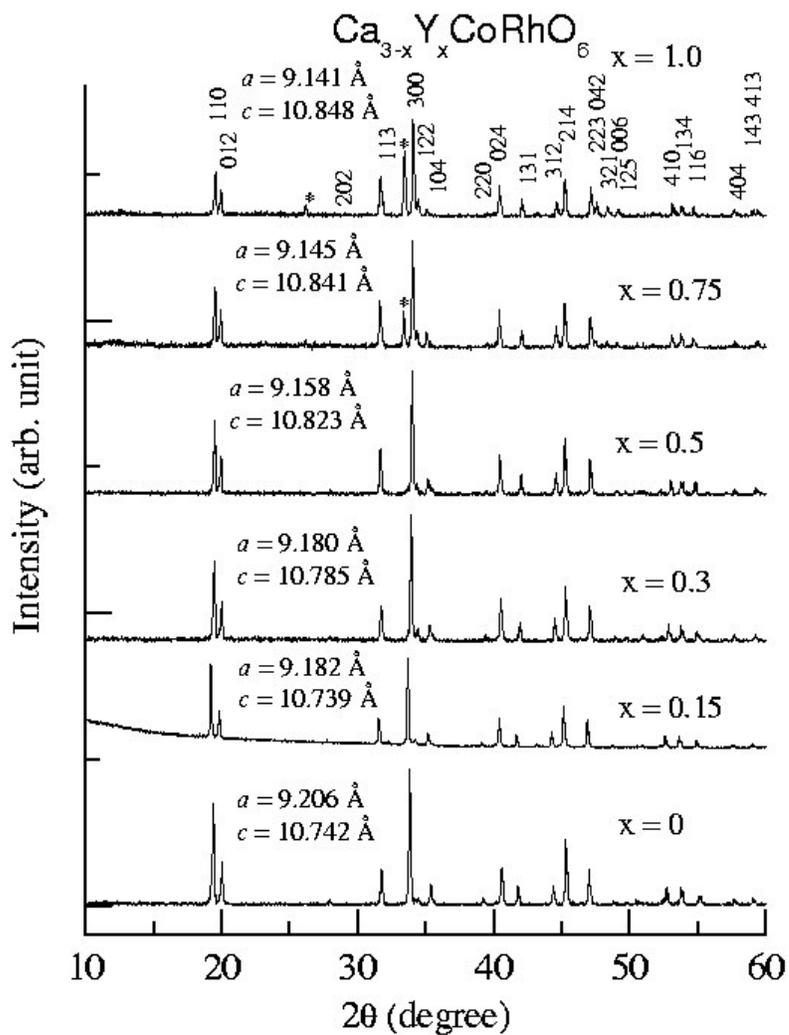

Figure 1: X-ray diffraction patterns (Cu $K_\alpha$) of solid solution, $Ca_{3-x}Y_xCoRhO_6$ (x= 0.0, 0.15, 0.3, 0.5, 0.75 and 1.0). The asterisks indicate extra phase.

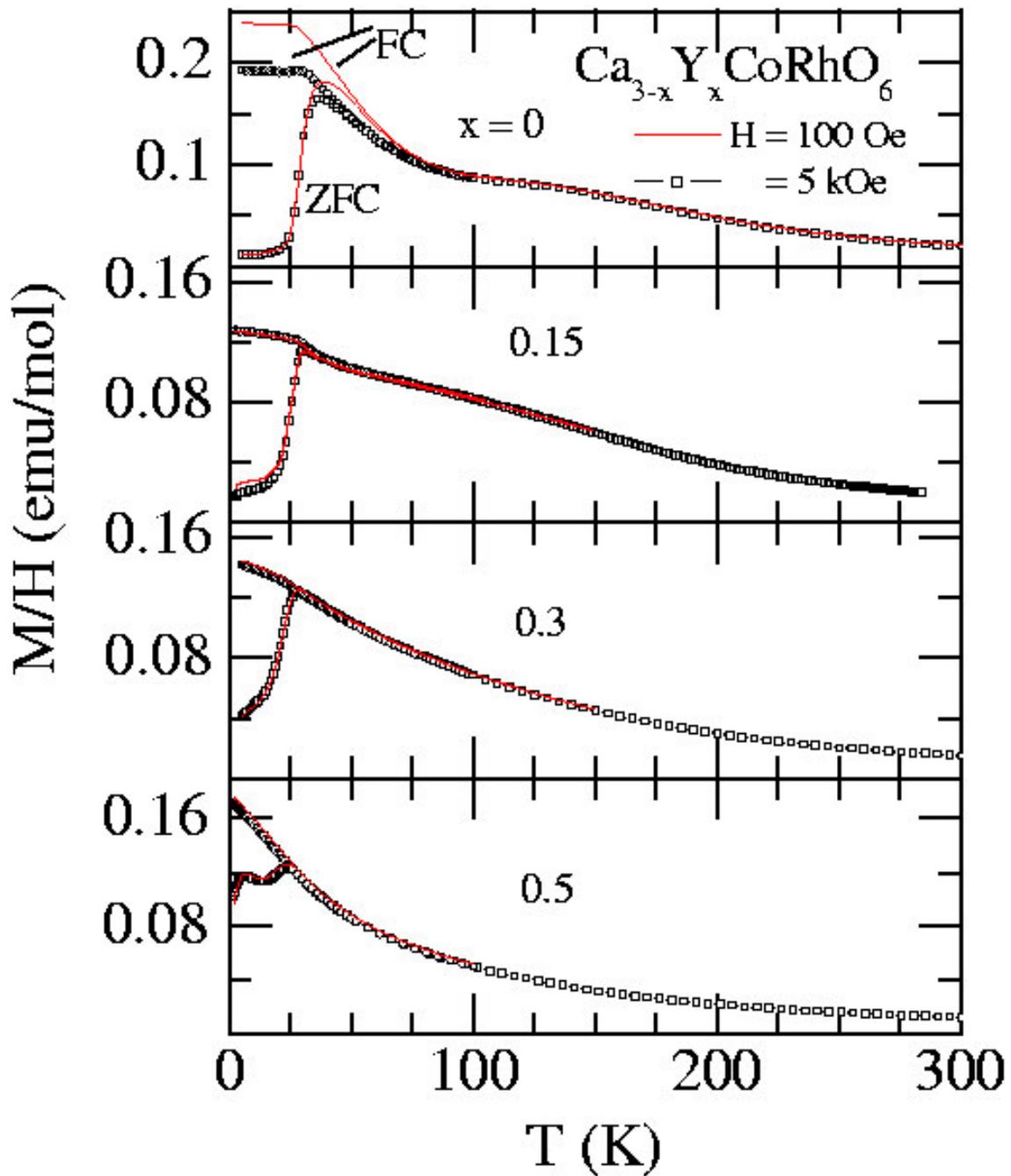

Figure 2: (color online) Dc magnetization divided by magnetic field as a function of temperature in the solid solution, $Ca_{3-x}Y_xCoRhO_6$, for the zero-field-cooled and field-cooled conditions of the specimens, in the presence of 100 Oe and 5 kOe. The curves for each composition (except for $x= 0.0$) overlap for both the fields for a given history of sample cooling.

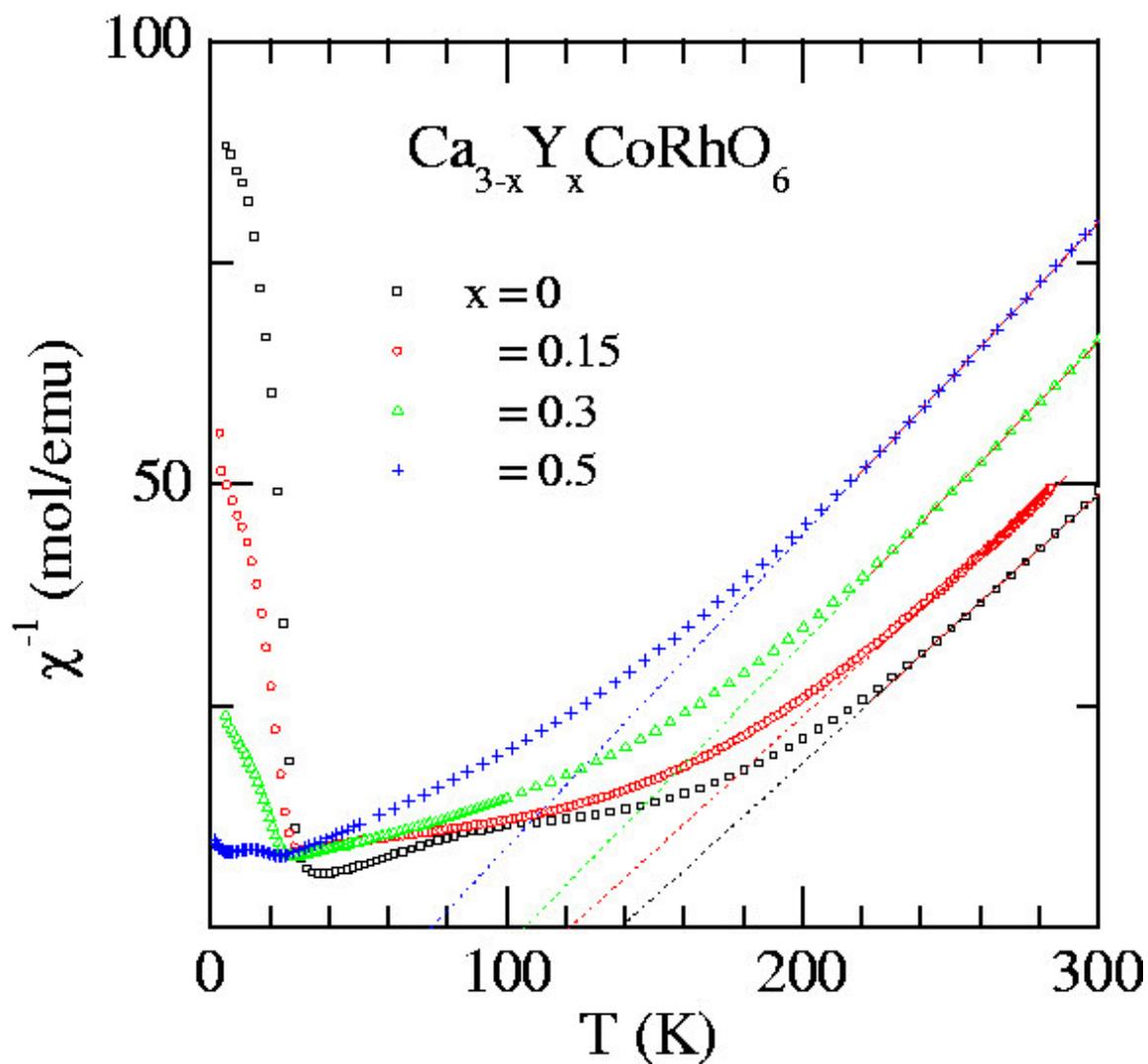

Figure 3: (color online) Inverse susceptibility as a function of temperature for Ca$_{3-x}$Y$_x$CoRhO$_6$ in the paramagnetic state. A discontinuous line is drawn through the high temperature linear region.

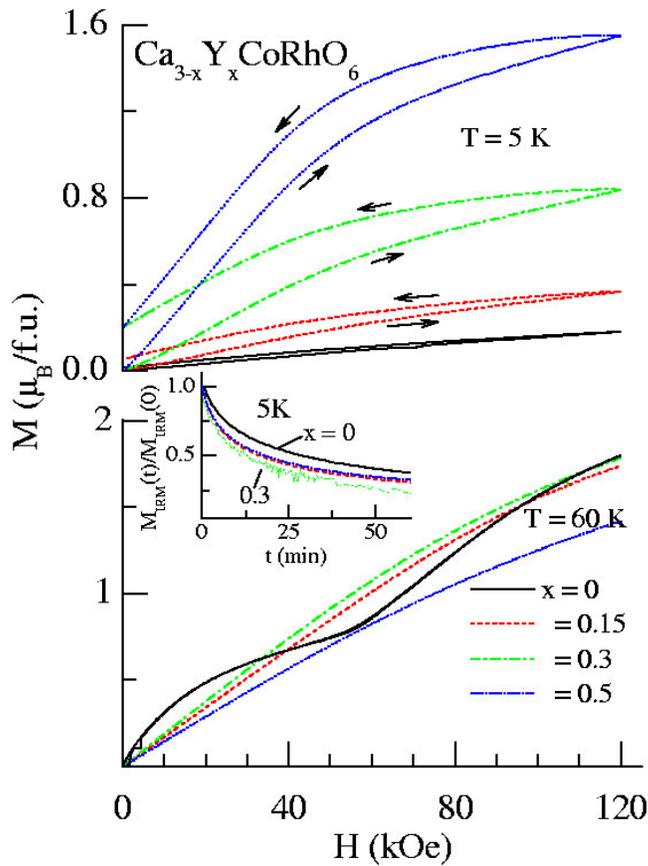

Figure 4: (color online) Isothermal magnetization at 5 and 60 K for $Ca_{3-x}Y_xCoRhO_6$ for the zero-field-cooled conditions of the specimens. Isothermal remanent magnetization data at 5 K are also shown in the insets; in this inset, the curves, for $x=$ 0.15 and 0.5 overlap.

Figure 5: (color online) Real ($\chi'$) and imaginary ($\chi''$) parts of ac susceptibility for $Ca_{3-x}Y_xCoRhO_6$ measured with different frequencies and an ac field 1 Oe, in zero and 30 kOe dc field. The arrows indicate the direction in which the curves move with increasing frequency ($\nu$). For the data for $x=$ 0.0 in the presence of 30 kOe, see Ref. 6. The lines through the data points serve as a guide to the eyes. For the sake of clarity, the data with too much of noise either are not joined by lines or have been omitted (for $x=$ 0.5, 30 kOe data) for the sake of clarity.

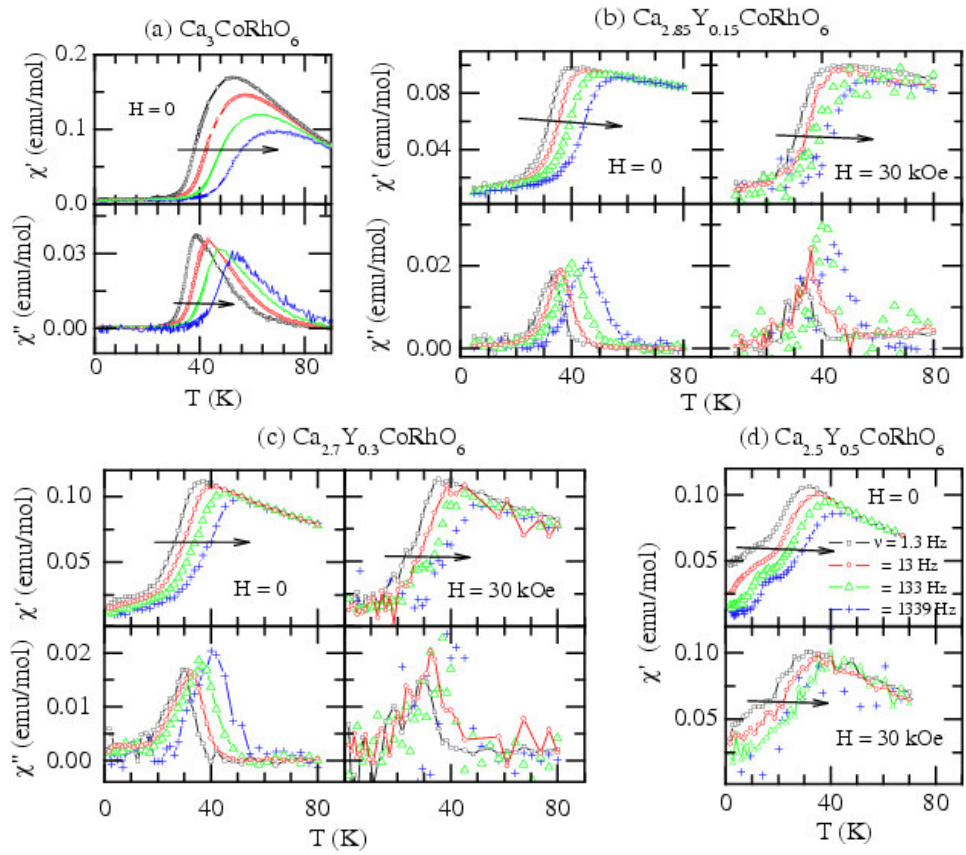